\documentclass[pra,twocolumn]{revtex4-1}

\usepackage{amsmath,amssymb, bbold, bbm, soul,bm}
\usepackage[pdftex]{graphicx}
\usepackage{epstopdf}
\usepackage{color}
\usepackage{algorithm}

\newcommand{\beq}{\begin{equation}}
\newcommand{\eeq}{\end{equation}}

\newcommand{\ket}[1]{| #1 \rangle}
\newcommand{\bra}[1]{\langle #1 |}

\newcommand{\matele}[3]{\langle #1 | #2 | #3 \rangle}

\newcommand{\by}[1]{#1 \times #1}
\newcommand{\iner}{{\rm In}}

\newcommand{\rk}{{\rm Rank}}
\newcommand{\tr}{{\rm Tr}}
\newcommand{\nn}{\nonumber\\}
\newcommand{\ii}{{\rm i}}
\newcommand{\comment}[1]{}

\begin{document}

\title{Strictly-complete measurements for bounded-rank quantum-state tomography}

\author{Charles H. Baldwin}
\email{baldwin4@unm.edu}
\author{Ivan H. Deutsch}
\email{ideutsch@unm.edu}
\author{Amir Kalev}
\email{amirk@unm.edu}
\affiliation{Center for Quantum Information and Control, MSC07--4220, University of New Mexico, Albuquerque, New Mexico 87131-0001, USA}
\date{\today}

\begin{abstract}
We consider the problem of quantum-state tomography under the assumption that the state is pure, and more generally that its rank is bounded by a given value $r$.  In this scenario two notions of informationally complete measurements emerge: rank-$r$ complete measurements and rank-$r$ strictly-complete measurements. Whereas in the first notion, a rank-$r$ state is uniquely identified from within the set of rank-$r$ states, in the second notion the same state is uniquely identified from within the set of all physical states, of any rank. We argue, therefore, that strictly-complete measurements are compatible with convex optimization, and we prove that they allow robust quantum state estimation in the presence of experimental noise. We also show that rank-$r$ strictly-complete measurements are as efficient as rank-$r$ complete measurements. We construct examples of strictly-complete measurements and give a complete description of their structure in the context of matrix completion. Moreover, we numerically show that a few random bases form such measurements.  We demonstrate the efficiency-robustness property for different strictly-complete measurements with numerical experiments. We thus conclude that only strictly-complete measurements are useful for practical tomography.
\end{abstract}

\maketitle

\section{Introduction} 
Quantum-state tomography (QST) is a standard tool used to characterize, validate, and verify the performance of quantum information processors.  Unfortunately, QST is a demanding experimental task, partly because the number of free parameters of an arbitrary quantum state scale quadratically with the dimension of the system. To overcome this difficulty, one can study QST protocols which include prior information about the system and effectively reduce the number of free parameters in the model. In this work we study QST under the prior information that the state of the system is close to a pure state, and more generally, that it is close to a bounded-rank state  (a density matrix with rank less than or equal to a given value). Indeed, in most quantum information processing applications the goal is not to manipulate arbitrary states, but to create and coherently evolve pure states. When the device is performing well, and there are only small errors, the quantum state produced will be close to a pure state, and the density matrix will have a dominant eigenvalue. One can use other techniques, e.g. randomized benchmarking~\cite{emerson05, knill08, magesan11}, to gain confidence that it is operating near this regime. This important prior information can be applied to significantly reduce the resources required for QST. We study different aspects of informational completeness that allow for efficient estimation in this scenario, and robust estimation in the face of noise or when the state is full rank, but still close to a bounded-rank state. 

Bounded-rank QST has been studied by a number of previous workers \cite{Flammia05,Finkelstein04,gross10,liu11,Heinosaari13,Chen13,Goyeneche14,Carmeli14,Kalev15,Carmeli15,Kech15,Ma16}, and has been shown to require less resources than general QST. One approach is based on the compressed sensing methodology~\cite{gross10,liu11}, where certain sets of randomly chosen measurements guarantee a robust estimation of low-rank states with high probability. Other schemes~\cite{Flammia05,Finkelstein04,Heinosaari13,Chen13,Goyeneche14,Carmeli15,Kech15,Ma16}, not related to compressed sensing, construct specific measurements that accomplish bounded-rank QST, and some of these protocols have been implemented experimentally~\cite{Goyeneche14,Hector15}. In addition, some general properties of such measurements have been derived~\cite{Carmeli14, Kalev15, Kech15}.

When considering bounded-rank QST a natural notion of informational completeness emerges~\cite{Heinosaari13}, referred to as {\em rank-$r$ completeness}. A measurement (a POVM) is rank-$r$ complete if the outcome probabilities uniquely distinguish a state with rank $\leq r$ from any other state with rank $\leq r$. A rigorous definition is given below.  The set of quantum states with rank $\leq r$, however, is not convex, and in general we cannot construct efficient estimators based on rank-$r$ complete measurements that will yield a reliable state reconstruction in the presence of experimental noise. This poses a concern for the practicality of such measurements for QST.

The purpose of this contribution is two fold: (i) We develop the significance of a different notion of informational completeness that we denote as {\em rank-$r$ strict-completeness}. We prove that strictly-complete measurements allow for robust estimation of bounded-rank states in the realistic case of noise and experimental imperfections by solving essentially any convex program. Because of this, strictly-complete measurements are crucial for the implementation of bounded-rank QST.  (ii) We study two different types of strictly-complete measurements and show that they require less resources than general QST. The first is a special type of measurement called ``element-probing'' POVM (EP-POVM). For example, the measurements proposed in Refs.~\cite{Flammia05,Goyeneche14} are EP-POVMs. In this context, the problem of QST translates to the problem of density matrix completion, where the goal is to recover the entire density matrix when only a few of its elements are given. The formalism we develop here entirely captures the underlying structure of all EP-POVMs and solves the problem of bounded-rank density matrix completion. The second type of strictly-complete POVM we study is the set of Haar-random bases. Based on numerical evidence we argue that measurement outcomes of a few random bases form a strictly-complete POVM and that the number of bases required to achieve strict completeness scales weakly with the dimension and rank.

The remainder of this article is organized as follows.  In Sec.~\ref{sec:Info Complete} we establish the different definitions of informational completeness and in Sec.~\ref{sec:Power of strict} we demonstrate the power of strictly-complete POVMs for practical tomography. We show how such POVMs allow us to employ convex optimization tools in quantum state estimators, and how the result is robust to experimental noise.  In Sec.~\ref{sec:constructions} we establish a complete theory of rank-$r$ complete and strictly-complete POVMs for the case of EP-POVMs and explore numerically how measurements in random orthogonal bases yield a strictly-complete POVM. We also demonstrate the robustness of strictly-complete measurements with numerical simulations of noisy measurements in Sec.~\ref{sec:numerics}. We summarize and conclude in Sec.~\ref{sec:conclusions}.

\section{Informational completeness in bounded-rank QST} \label{sec:Info Complete}
\begin{figure}[t]
\centering
\includegraphics[width=\linewidth]{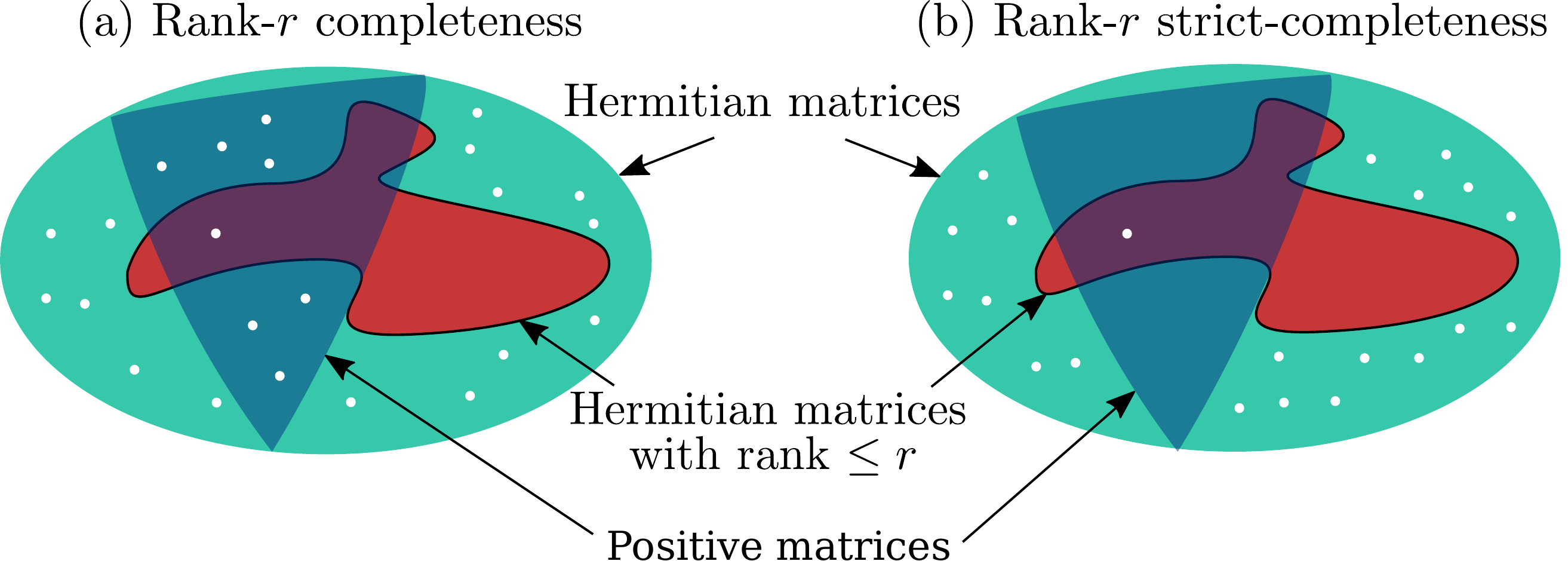}
\caption{{\bf Various notions of completeness in bounded-rank QST.} The white dots represent Hermitian matrices, positive or not, that are consistent with the (noiseless) measurement record.  {\bf (a) Rank-$r$ completeness.} The measurement record, distinguishes the rank $\leq r$ state from any other rank $\leq r$ PSD matrix. However, there generally will be infinitely many other states,  with rank greater than $r$, that are consistent with the measurement record.  {\bf (b)  Rank-$r$ strict-completeness.}  The measurement record  distinguishes the rank $\leq r$ state from any other PSD matrix. Thus it is unique in the convex set of PSD matrices.}
\label{fig:illustration}
\end{figure}
QST has two basic ingredients, states and measurements, so it is important to define these precisely.  A quantum state in a $d$-dimensional Hilbert space, ${\mathcal H}_d$, is a density matrix, $\rho$, that is positive semidefinite (PSD) and normalized to unit trace. A quantum measurement with $m$ possible outcomes (events) is defined by a positive-operator valued measure (POVM) with $m$ elements,  ${\cal E}=\{E_\mu:E_\mu \geq 0,\, \sum_{\mu=1}^m E_\mu=\mathbb{1}\}$.  A POVM then has an associated map ${\cal M}_{\cal E}[\cdot]=[\tr(E_1\cdot),\ldots,\tr(E_m\cdot)]$. This map acts on a quantum state $\rho$ to return a vector of probabilities, ${\bm p}\equiv{\cal M}_{\cal E}[\rho]=[\tr(E_1 \rho),\ldots,\tr(E_m \rho)]$, which we refer to this as the ``measurement vector." A POVM is {\em fully informationally complete} if the measurement vector, ${\bm p}$, distinguishes the state $\rho$ from all other states. A fully informationally complete POVM must have $d^2$ linearly independent elements.

We commonly think of POVMs as acting on quantum states, but mathematically we can apply them, more generally, on PSD matrices. In this work we discuss POVMs acting on PSD matrices as it highlights the fact that our definitions and results are independent of the trace constraint of quantum states, and only depend on the positivity property. To accomplish this we treat the map, ${\cal M}_{\cal E}[\cdot]$ defined by a POVM ${\cal E}$,  more generally as a map between the space of PSD matrices and the real vector space, ${\mathbb R}^m$. Particularly, the action of this map on $X \geq 0$ is given as ${\cal M}_{\cal E}[X] = {\bf y}$ where $y_\mu\geq0$ and $\sum_{\mu=1}^m y_\mu=\tr (X)$. The second expression shows that since, by definition, the POVM elements sum to the identity, the POVM always measures the trace of the matrix. It is also useful to define the kernel of a POVM, ${\rm Ker}({\mathcal E})\equiv \{X:{\cal M}_{\cal E}[X]={\bf 0}\}$. Since the POVM elements sum to the identity matrix, we immediately obtain that every $X\in{\rm Ker}({\mathcal E})$ is traceless, $\tr(X)=0$.

In bounded-rank QST, rank-$r$ completeness is a natural concept~\cite{Heinosaari13, Carmeli14, Kalev15}, which we define here in terms of PSD matrices.\vspace{0.2cm}\\
\noindent {\bf Definition~1:} Let  ${\cal S}_r=\{X:X\geq0,\rk X\leq r\}$ be the set of $\by{d}$ PSD matrices with rank $\leq r$.  A POVM is said to be {\em rank-$r$ complete} if  
\begin{equation} 
\label{restricted_definition}
\forall \, X_1, X_2 \in {\cal S}_r, X_1\neq X_2, \, {\cal M}_{\cal E}[X_1]\neq{\cal M}_{\cal E}[X_2],
\end{equation}
except for possibly a set of rank-$r$ states that are dense on a set of measure zero, called the ``failure set.''
\vspace{0.2cm}\\
In the context of QST, the measurement probabilities of a rank-$r$ complete POVM uniquely identify the rank $\leq r$ state from within the set of all PSD matrices with rank $\leq r$, ${\cal S}_r$. Figure~\ref{fig:illustration}a illustrates the notion of rank-$r$ completeness. The measurement probabilities cannot uniquely identify states in this way if they lie in the failure set, as was considered in~\cite{Flammia05,Goyeneche14}. The chances of randomly hitting a state in that set, however, is vanishingly small. While in Refs.~\cite{Heinosaari13, Carmeli14, Kalev15} the definition of rank-$r$ completeness does not include a failure set, here we chose to include it following~\cite{Flammia05}. We comment on the implications and structure of the failure set for practical tomography in Sec.~\ref{sec:constructions}.

Using ${\rm Ker}({\mathcal E})$ we arrive at an alternative, equivalent, definition for rank-$r$ complete:  A POVM ${\mathcal E}$ is rank-$r$ complete if $\forall \, X_1,X_2 \in {\cal S}_r$, with $\, X_1\neq X_2$, the difference $\Delta= X_1-X_2$ is not in the kernel of ${\mathcal E}$, i.e., $\exists \, E_\mu$ such that $\tr(E_\mu \Delta)\neq0$. Carmeli~{\em et al.}~\cite{Carmeli14} showed that a necessary and sufficient condition for a POVM to be rank-$r$ complete is that every nonzero $\Delta\in{\rm Ker}({\mathcal E})$ has $\max(n_{-},n_{+})\geq r+1$, where $n_{+}$ and $n_{-}$ are the number of strictly positive and strictly negative eigenvalues of a matrix, respectively. This condition was derived for PSD matrices with rank $\leq r$. If we exclude the positivity property, and only consider the rank property, we obtain a sufficient condition:  a POVM is rank-$r$ complete if every nonzero $\Delta\in{\rm Ker}({\mathcal E})$ has $\rk(\Delta)\geq 2r+1$. This sufficient condition applies to all Hermitian matrices with rank $\leq r$, PSD or otherwise.  Using the sufficient condition alone, it was shown~\cite{Heinosaari13} that the expectation values of particular $4r(d-r)$ observables corresponds to rank-$r$ complete measurement, and moreover~\cite{Kech15} that a measurement of $4r\lceil\frac{d-r}{d-1}\rceil$ random orthonormal bases is rank-$r$ complete. 

The definition of rank-$r$ complete POVMs guarantees the uniqueness of the reconstructed state in the set ${\cal S}_r$, but it does not say anything about higher-rank states. There may be other density matrices, with rank greater than $r$ that are consistent with the measurement probabilities. Since ${\cal S}_r$ is a nonconvex set it may be difficult to differentiate between the unique rank-$r$ density matrix and these higher-rank states, particularly in the presence of noise or other experimental imperfections. To overcome this difficulty, we consider a ``stricter" type of POVM which excludes these higher-rank states.  This motivates the following definition~\cite{Chen13,Carmeli14,Kalev15}\vspace{0.2cm}:\\
\noindent {\bf Definition~2:} Let  ${\cal S}=\{X:X\geq0\}$ be the set of $\by{d}$ PSD matrices. A POVM is said to be {\em rank-$r$ strictly-complete} if
\begin{equation} 
\label{strictly_definition}
 \forall \, X_1 \in {\cal S}_r, \, \forall \, X_2 \in {\cal S}, X_1\neq X_2, \, {\cal M}_{\cal E}[X_1]\neq{\cal M}_{\cal E}[X_2],
\end{equation}
except for possibly a set of rank-$r$ states that are dense on a set of measure zero, called the ``failure set.''
\vspace{0.2cm}\\
The implication for QST is that when the rank of the state being measured is promised to be smaller than or equal to $r$, the measurement probabilities of a rank-$r$ strictly-complete POVM distinguish this state from any other PSD matrix, of {\em any} rank (except on the failure set). Figure~\ref{fig:illustration}b illustrates the notion of rank-$r$ strict-completeness. 

Carmeli~{\em et al.}~\cite{Carmeli14} showed that a POVM is rank-$r$ strictly-complete if, and only if, every nonzero $X\in{\rm Ker}({\mathcal E})$ has $\min(n_{-},n_{+})\geq r+1$.  This condition relies on the PSD property of the matrices. To date, there are only a few known POVMs that are rank-$r$ strictly-complete~\cite{Chen13,Ma16}. In Sec.~\ref{sec:constructions}, we present new strictly-complete POVMs with ${\cal O}(rd)$ elements.

In contrast to the notion of rank-$r$ completeness, which can be defined generally for bounded-rank Hermitian matrices, the definition of strict-completeness is nontrivial only when applied to PSD matrices, and therefore, in particular to quantum states. To see this, let us apply the definition of strict-completeness for bounded-rank Hermitian matrices, ignoring positivity. Let $R$ be a Hermitian matrix with $\rk(R)\leq r$.  To be (nontrivially) strictly-complete the POVM should be able to distinguish $R$ from any Hermitian matrix, of any rank, with less than $d^2$ linearly independent POVM elements. (If the POVM has $d^2$ linearly independent POVM elements, it is fully informationally complete and can distinguish any Hermitian matrix from any other.) However, a POVM with less than $d^2$ linearly independent elements necessarily has infinitely many Hermitian matrices, with rank $> r$, which produce the same noiseless measurement vector as $R$. Therefore, without positivity, we cannot define strict-completeness with less than $d^2$ linearly independent elements. On the other hand, if we impose positivity, as we will shortly see, there exists POVMs that are rank-$r$ strictly-complete and have fewer than $d^2$ elements.

\section{The power of strictly-complete measurements} ~\label{sec:Power of strict}
The usefulness of strictly-complete measurements become evident when we consider implementations in  a realistic experimental context. It is essential that the estimation protocol be robust to noise and other imperfections. Thus, any realistic estimation procedure should allow one to find the closest estimate (by some appropriate measure of distance) given a bound on the noise. We can address this by convex optimization, whereby the estimate is found by minimizing a convex function over a convex set. Since rank-$r$ strictly-complete POVMs uniquely identify a rank-$r$ quantum state from within the {\em convex} set of PSD matrices, the data obtained from measurements defined by these POVMs fits the convex optimization paradigm. On the other hand, rank-$r$ complete measurements uniquely identify the rank-$r$ state only from within the {\em nonconvex} set of rank-$r$ PSD matrices, and therefore are not compatible to use with convex optimization. 

This is formalized in the following corollary for the noiseless measurement case:
\vspace{0.1cm}\\
\noindent {\bf Corollary~1 (uniqueness):} Let $\rho_0$ be a quantum state with rank $\leq r$, and let ${\bm p}= {\cal M}_{\cal E}[\rho_0]$ be the corresponding measurement vector of a rank-$r$ strictly-complete POVM ${\cal E}$. Then, the solution to
\begin{equation}
\label{general_positive_CS}
\hat{X} = \arg\min_X \mathcal{C}(X)\;\; {\rm s.t.}\; {\cal M}[X]=\bm{p}\, \,  {\rm and} \, \,  X \geq 0,
\end{equation}
or to
\begin{equation}
\label{general_norm_positive_CS}
\hat{X} = \arg\min_X \Vert{\cal M}[X]-\bm{p}\Vert\;\; {\rm s.t.}\; X \geq 0,
\end{equation}
where $\mathcal{C}(X)$ is a any convex function of $X$, and $\Vert\cdot\Vert$ is any norm function, is unique:  $\hat{X} = \rho_0$.\vspace{0.1cm}\\
\noindent {\em Proof:} This is a direct corollary of the definition of strict-completeness. Since, by definition, the probabilities of rank-$r$ strictly-complete POVM uniquely determine $\rho_0$ from within the set of all PSD matrices, its reconstruction becomes a feasibility problem over the convex set $\{{\cal M}[X]=\bm{p},X\geq0\}$,
\begin{equation} \label{feasibility}
{\rm find}\; X\;\; {\rm s.t.}\; {\cal M}[X]=\bm{p}\, \,  {\rm and} \, \, X \geq 0.
\end{equation}
The solution for this feasibility problem is $\rho_0$ uniquely. Therefore, any optimization program, and particularly an efficient convex optimization program that looks for the solution within the feasible set, is guaranteed to find $\rho_0$.\hfill $\square$

Corollary~1 was proved in Ref.~\cite{Kech15b} for the particular choice $\mathcal{C}(X)=\tr(X)$, and also in the context of compressed sensing measurements in Ref.~\cite{Kalev15}. Note, while one can also include a trace constraint $\tr(X) = t$, in this noiseless case, Eqs.~\eqref{general_positive_CS} and~\eqref{general_norm_positive_CS}, this is redundant since any POVM ``measures'' the trace of a matrix.  Thus, if we have prior information that $\tr(X)=t$, then the feasible set in Eq.~\eqref{feasibility} is equal to the set $\{ X \, | \,{\cal M}[X]=\bm{p}, \, X \geq 0,\,  {\rm and} \, \tr(X) = t \}$.  In particular, in the context of QST with idealized noiseless data, the constraint $\tr\rho=1$ would be redundant; the reconstructed state would necessarily be properly normalized.

This corollary implies that strictly-complete POVMs allow for the reconstruction of bounded-rank states via convex optimization even though the set of bounded-rank states is nonconvex. Moreover, all convex programs over the feasible solution set, i.e., of the form of Eqs.~\eqref{general_positive_CS} and~\eqref{general_norm_positive_CS}, are equivalent for this task. For example, this result applies to maximum-(log)likelihood estimation~\cite{Hradil97} where $\mathcal{C}(\rho) =-\log( \prod_{\mu}\tr(E_\mu\rho)^{p_\mu})$.  Corollary~1 does not apply for states in the POVM's failure set, if such set exists. 

It is also essential that the estimation protocol be robust to noise and other imperfections. In any real experiment the measurement vector necessarily contains noise due to finite statistics and systematic errors.  Moreover, any real state assignment should have full rank, and the assumption that the state has rank $\leq r$ is only an approximation.  Producing a robust estimate in this case with rank-$r$ complete measurements is generally a hard problem since the set of bounded-rank states in not convex.  Strict-completeness, however, together with the convergence properties of convex programs, ensure a robust state estimation in realistic experimental scenarios. This is the main advantage of strictly-complete measurements and is formalized in the following corollary.  \vspace{0.1cm}\\
\noindent {\bf Corollary~2 (robustness):} Let $\sigma$ be the state of the system, and let ${\bm f}= {\cal M}_{\cal E}[\sigma]+{\bm e}$ be the (noisy) measurement vector of a rank-$r$ strictly-complete POVM ${\cal E}$, such that $\Vert{\bm e}\Vert\leq\epsilon$. If  $\Vert{\bm f}-{\cal M}_{\cal E}[\rho_0]\Vert\leq \epsilon$ for some quantum state $\rho_0$ with $\rk(\rho_0)\leq r$, then the solution to
\begin{equation}
\label{general_positive_CS_noisy}
\hat{X} = \arg\min_X \mathcal{C}(X)\;\; {\rm s.t.}\; \Vert{\cal M}[X]-\bm{f}\Vert\leq\epsilon\, \,  {\rm and} \, \,  X \geq 0,
\end{equation}
or to
\begin{equation}
\label{general_norm_positive_CS_noisy}
\hat{X} = \arg\min_X \Vert{\cal M}[X]-\bm{f}\Vert\;\; {\rm s.t.}\; X \geq 0,
\end{equation}
where $\mathcal{C}(X)$ is a any convex function of $X$, is robust:  $\Vert\hat{X} - \rho_0\Vert\leq C_{\cal E}\epsilon$, and $\Vert\hat{X} - \sigma\Vert\leq 2C_{\cal E}\epsilon$, where $\Vert\cdot\Vert$ is any $p$-norm, and $C_{\cal E}$ is a constant which depends only on the POVM.\vspace{0.1cm}\\
\\
\noindent The proof, given in Appendix~\ref{app:proof}, is derived from Lemma~V.5 of Ref.~\cite{Kech15b}  where it was proved for the  particular choice $\mathcal{C}(X)=\tr(X)$. In Ref.~\cite{Kalev15} this was also studied in the context of compressed sensing measurements. This corollary assures that if the state of the system is close to a bounded-rank density matrix and is measured with strictly-complete measurements, then it can be robustly estimated with any convex program, constrained to the set of PSD matrices. In particular, it implies that all convex estimators perform qualitatively the same for low-rank state estimation. This may be advantageous especially when considering estimation of high-dimensional quantum states. 

As in the noiseless case, the trace constraint is not necessary for Corollary~2, and in fact leaving it out allows us to make different choices for $\mathcal{C}(X)$, as was done in Ref.~\cite{Kech15b}. However, for a noisy measurement vector, the estimated matrix $\hat{X}$ is generally not normalized, $\tr\hat{X}\neq1$. The final estimation of the state is then given by $\hat{\rho} = \hat{X}/\tr(\hat{X})$. In principle, we can consider a different version of Eqs.~\eqref{general_positive_CS_noisy} and~\eqref{general_norm_positive_CS_noisy} where we include the trace constraint, and this may have implications for the issue of ``bias" in the estimator. This will be studied in more details elsewhere.

\section{Building strictly-complete measurements}\label{sec:constructions}
So far, we have shown that strictly-complete measurements are advantageous because of their compatibility with convex optimization (Corollary~1) and their robustness to statistical noise and to state preparation errors (Corollary~2).  We have not, however, discussed how to find such measurements or the resources required to implement them. In this section we answer these questions with two different approaches. First, we describe a general framework that can be used to construct strictly-complete EP-POVMs, and we explicitly construct two examples of such POVMs in Appendix~\ref{app:construction}. Second, we numerically study the number of random bases that corresponds to strictly-complete POVMs for certain states rank and dimension. In both cases we find that the number of POVM elements required is ${\cal O}(rd)$, implying that strictly-complete measurements can be implemented efficiently.

\subsection{Element-probing POVMs}\label{ssec:ep}
EP-POVMs are special types of POVMs where the measurement probabilities determine a subset of the total $d^2$ matrix elements, referred to as the ``measured elements.'' With EP-POVMs, the task of QST is to reconstruct the remaining (unmeasured) density matrix elements from the measured elements, and thus, in this case, QST is equivalent to the task of density matrix completion. Similar work was carried out in Ref.~\cite{Bakonyi11} to study the problem of PSD matrix completion.

Examples of EP-POVMs were studied by Flammia~{\em et al.}~\cite{Flammia05}, and more recently, by Goyeneche~{\em et al.}~\cite{Goyeneche14}, and shown to be rank-1 complete. We briefly review them here since we use them as canonical examples for the framework we develop. Flammia~{\em et al.}~\cite{Flammia05} introduced the following POVM,
\begin{align}\label{psi-complete pure}
&E_0=a\ket{0}\bra{0},\nn
&E_n=b(\mathbb{1}+\ket{0}\bra{n}+\ket{n}\bra{0}),\; \;n=1,\ldots,d-1,\nn
&\widetilde{E}_n=b(\mathbb{1}-\ii\ket{0}\bra{n}+\ii\ket{n}\bra{0}),\; \;n=1,\ldots,d-1,\nn
&E_{2d}=\mathbb{1}-\Bigl[E_0 +\sum_{n=1}^{d-1}(E_n+\widetilde{E}_n)\Bigr],
\end{align}
with $a$ and $b$ chosen such that $E_{2d}\geq0$.  They showed that the measurement probabilities $p_\mu=\bra{\psi}E_\mu\ket{\psi}$ and $\tilde{p}_\mu=\bra{\psi}\widetilde{E}_\mu\ket{\psi}$ can be used to reconstruct any $d$-dimensional pure state $\ket{\psi}=\sum_{k=0}^{d-1}c_k\ket{k}$, as long as  $c_0\neq0$. Under the assumption, $c_0>0$, we find that  $c_0=\sqrt{p_0}/a$. The real and imaginary parts of $c_n$, $n=1,\ldots, d-1$, are found through the relations $\Re(c_n)=\frac1{2c_0}(\frac{p_n}{b}-1)$ and $\Im(c_n)=\frac1{2c_0}(\frac{\tilde{p}_n}{b}-1)$, respectively. The POVM in Eq.~\eqref{psi-complete pure} is in fact an EP-POVM where the measured elements are the first row and column of the density matrix. The probability $p_0=\tr(E_0\rho)$ can be used to algebraically reconstruct the element $\rho_{0,0}=\matele{0}{\rho}{0}$, and the probabilities $p_n=\tr(E_n\rho)$ and $\tilde{p}_n=\tr(\widetilde{E}_n\rho)$ can be used to reconstruct the elements $\rho_{n,0}=\matele{n}{\rho}{0}$ and $\rho_{0,n}=\matele{0}{\rho}{n}$, respectively. Further details of this construction are given in Appendix~\ref{app:examples_full}.

A second EP-POVM that is rank-1 complete was studied by Goyeneche~{\em et al.}~\cite{Goyeneche14}. In this scheme four specific orthogonal bases are measured, 
\begin{align}\label{4gmb}
\mathbbm{B}_{1} &=\Bigl\{ \frac{\ket{0}\pm\ket{1}}{\sqrt2}, \frac{\ket{2}\pm\ket{3}}{\sqrt2}, \ldots, \frac{\ket{d-2}\pm\ket{d-1}}{\sqrt2}\Bigr\}, \nonumber \\
\mathbbm{B}_{2} &=\Bigl\{ \frac{\ket{1}\pm\ket{2}}{\sqrt2}, \frac{\ket{3}\pm\ket{4}}{\sqrt2}, \ldots, \frac{\ket{d-1}\pm\ket{0}}{\sqrt2}\Bigr\}, \nonumber \\
\mathbbm{B}_{3} &=\Bigl\{ \frac{\ket{0}\pm\ii\ket{1}}{\sqrt2}, \frac{\ket{2}\pm\ii\ket{3}}{\sqrt2}, \ldots, \frac{\ket{d-2}\pm\ii\ket{d-1}}{\sqrt2}\Bigr\}, \nonumber \\
\mathbbm{B}_{4} &=\Bigl\{ \frac{\ket{1}\pm\ii\ket{2}}{\sqrt2}, \frac{\ket{3}\pm\ii\ket{4}}{\sqrt2}, \ldots, \frac{\ket{d-1}\pm\ii\ket{0}}{\sqrt2}\Bigr\}.
\end{align}
Goyeneche~{\em et al.}~\cite{Goyeneche14} outlined a procedure to reconstruct the pure state amplitudes but we omit it here for brevity. Similar to the POVM in Eq.~\eqref{psi-complete pure}, the procedure fails when certain state-vector amplitudes vanish. More details are given in Appendix~\ref{app:construction}. This POVM is an EP-POVM as well. Here, the measured elements are the elements on the first diagonals (the diagonals above and below the principal diagonal) of the density matrix. Denoting $p_{j}^{\pm}=\frac{1}{2}(\bra{j}\pm\bra{j+1})\rho(\ket{j}\pm\ket{j+1})$, and $p_{j}^{\pm\ii}=\frac{1}{2}(\bra{j}\mp\ii\bra{j+1})\rho(\ket{j}\pm\ii\ket{j+1})$, we obtain, $\rho_{j,j+1}{=}\frac{1}{2}[(p_{j}^{+}-p_{j}^{-})+\ii (p_{j}^{+\ii}-p_{j}^{-\ii})]$ for $j = 0,\ldots,d-1$, and addition of indices is taken modulo $d$. Goyeneche~{\em et al.}~\cite{Goyeneche14} also considered a protocol for measuring pure states by adaptively measuring five bases. In Appendix~\ref{app:examples_bases} we consider a related protocol with five-bases but without adaptation.

By their design, these two POVMs are rank-1 complete but are they rank-1 strictly-complete? Currently there is no unified and simple description of the underlying structure of EP-POVMs that allow for pure-state, and more generally bounded-rank state, identification. Moreover, due to the positivity constraint, it is generally difficult to determine if a EP-POVM is strictly-complete.  We address this issue by developing a framework that assess the completeness of EP-POVMs and explicitly deals with the positivity constraint . 

Our technique to determine the informational completeness of an EP-POVM relies on properties of the Schur complement and matrix inertia~\cite{Haynsworth68,Zhang11}. Consider a block-partitioned $\by{k}$ Hermitian matrix $M$,
\begin{equation} \label{block_mat}
M = 
\begin{pmatrix}
{A} & {B^{\dagger}} \\
{B} &{C} 
\end{pmatrix},
\end{equation}
where $A$ is a $\by{r}$ Hermitian matrix, and the size of ${B^{\dagger}}$, $B$ and $C$ is determined accordingly.
The Schur complement of $M$ with respect to $A$, assuming $A$ is nonsingular, is defined by
\begin{equation}
M/A \equiv C - B A^{-1} B^{\dagger}.
\end{equation}
The inertia of a Hermitian matrix is the ordered triple of the number of negative, zero, and positive eigenvalues, ${\rm In}(M)=(n_-, n_0, n_+)$, respectively. 

We will use the Haynsworth inertia additivity formula, which relates the inertia of $M$ to that of $A$ and of $M/A$~\cite{Haynsworth68},
\begin{equation} \label{Schur_iner}
\iner(M) = \iner(A)+\iner(M/A),
\end{equation}
A corollary of the inertia formula is the rank additivity property,
\begin{equation} \label{Schur_rank}
\rk(M) = \rk(A) + \rk(M/A).
\end{equation}
With these relations we can determine the informational completeness of any EP-POVM.

As an instructive example, we use these relations in an alternative proof that the POVM in Eq.~\eqref{psi-complete pure} is rank-1 complete without referring to the pure-state amplitudes. The POVM in Eq.~\eqref{psi-complete pure} is an EP-POVM, where the measured elements are $\rho_{0,0}$, $\rho_{n,0}$ and $\rho_{0,n}$ for $n=1,\ldots,d-1$. Supposing that $\rho_{0,0}>0$ and labeling the unmeasured $\by{(d-1)}$ block of the density matrix by $C$, we write
\begin{equation} \label{block_rho}
\rho=  \left(
    \begin{array}{cccc}
{\rho_{0,0}} &  {\rho_{0,1}}& \cdots &{\rho_{0,d-1}}\\
\cline{2-4}\multicolumn{1}{c|}{\rho_{1,0}}&
      {} &{}& \multicolumn{1}{c|}{} \\
      \multicolumn{1}{c|}{\vdots}&
      {} &{\;\;\Large\textit{C}}& \multicolumn{1}{c|}{}\\
\multicolumn{1}{c|}{\rho_{d-1,0}}&
      {} &{}& \multicolumn{1}{c|}{}\\\cline{2-4}
    \end{array}
    \right)
\end{equation}
Clearly, Eq.~\eqref{block_rho} has the same form as Eq.~\eqref{block_mat}, such that $M = \rho$, $A = \rho_{0,0}$, $B^{\dagger}=({\rho_{0,1}}\cdots {\rho_{0,d-1}})$, and $B = ({\rho_{0,1}}\cdots {\rho_{0,d-1}})^{\dagger}$. Assume $\rho$ is a pure state so $\rk(\rho)=1$. By applying Eq.~\eqref{Schur_rank} and noting that $\rk(A)=1$, we obtain $\rk(\rho/A)=0$. This implies that $\rho/A=C - B A^{-1} B^{\dagger}=0$, or equivalently, that $C= B A^{-1} B^{\dagger}= \rho_{0,0}^{-1}B B^{\dagger}$. Therefore, by measuring every element of $A$, $B$ (and thus of ${B^{\dagger}}$), the rank additivity property allows us to algebraically reconstruct $C$ uniquely without measuring it directly. Thus, the entire density matrix is determined by measuring its first row and column. Since we used the assumption that $\rk(\rho){=}1$, the reconstructed state is unique to the set ${\cal S}_1$, and the POVM is rank-1 complete. 

This algebraic reconstruction of the rank-$1$ density matrix works as long as $\rho_{0,0}\neq0$. When $\rho_{0,0}=0$, the Schur complement is not defined, and Eq.~\eqref{Schur_rank} does not apply. This, however, only happens on a set of states of measure zero (the failure set), i.e. the set of states where $\rho_{0,0} = 0$ exactly. It is exactly the same set found by Flammia~{\em et al.}~\cite{Flammia05}.

The above technique can be used to determine if any EP-POVM is rank-$r$ complete for a state $\rho\in{\cal S}_r$. In general, the structure of the measured elements will not be as convenient as the example considered above. Our approach is to study $\by{k}$ principle submatrices of $\rho$ such that $k > r$. Since $\rho$ is a rank-$r$ matrix, it has at least one nonsingular $\by{r}$ principal submatrix~\cite{footnote_rank}. Assume for now that a given $\by{k}$ principal submatrix, $M$, contains a nonsingular $\by{r}$ principle submatrix $A$. From Eq.~\eqref{Schur_rank}, since $\rk(M) = \rk(A) = r$, $\rk(M/A) = 0$, and therefore, $C = B A^{-1}B^{\dagger}$. This equation motivates our choice of $M$. If the measured elements make up $A$ and $B$ (and therefore $B^{\dagger}$) then we can solve for $C$ and we have fully characterized $\rho$ on the subspace defined by $M$. We refer to block-matrices in this form as a principal submatrix in the canonical form. In practice, the measured elements only need to be related to canonical form by a unitary transformation. In Appendix~\ref{app:examples_bases} we discuss such an example where the transformation is done by interchanging columns and corresponding rows. In general, an EP-POVM may measure multiple subspaces, $M_i$, and we can reconstruct $\rho$ only when the corresponding $A_i$, $B_i$, $C_i$ cover all elements of $\rho$~\cite{footnote_tr}. We label the set of all principle submatrices that are used to construct $\rho$ by ${\cal M} = \{M_i\}$. Since we can reconstruct a unique state within the set of $\mathcal{S}_r$ this is then a general description of a rank-$r$ complete EP-POVM.  The failure set, in which the measurement fails to reconstruct $\rho$, corresponds to the set of states that are singular on any of the $A_i$ subspaces. 

The structure defined above also allows us to to prove that under certain conditions, when we include the positivity constraint, a given EP-POVMs is strictly-informationally complete.  As an example, consider the rank-1 complete POVM in Eq.~\eqref{psi-complete pure}. Since $\rho/A=0$, by applying the inertia additivity formula to $\rho$ we obtain 
\begin{equation}
\iner(\rho) = \iner(A)+\iner(\rho/A)=\iner(A).
\end{equation}
This implies that $A$ is a PSD matrix. For the POVM in Eq.~\eqref{psi-complete pure}, $A=\rho_{0,0}$, so this equation is a re-derivation of the trivial condition $\rho_{0,0}\geq0$. Let us assume that the POVM is not rank-$1$ strictly-complete. If so, there must exist a PSD matrix, $\sigma\geq0$, with $\rk(\sigma)>1$, that has the same measurement vector and thus measured elements as $\rho$, but different unmeasured elements. We define this difference by $V\neq0$, and write
\begin{equation} \label{block_mat_sigma}
\sigma=  \left(
    \begin{array}{cccc}
{\rho_{0,0}} &  {\rho_{0,1}}& \cdots &{\rho_{0,d-1}}\\
\cline{2-4}\multicolumn{1}{c|}{\rho_{1,0}}&
      {} &{}& \multicolumn{1}{c|}{} \\
      \multicolumn{1}{c|}{\vdots}&
      {} &{\;\;\Large{\textit{C}}+\!\Large{\textit{V}}}& \multicolumn{1}{c|}{}\\
\multicolumn{1}{c|}{\rho_{d-1,0}}&
      {} &{}& \multicolumn{1}{c|}{}\\\cline{2-4}
    \end{array}
    \right)=\rho+\begin{pmatrix}
{0} &  {\bf 0}\\
{\bf 0} & V
\end{pmatrix}.
\end{equation}
Since $\sigma$ and $\rho$ have the same measurement vector, for all $\mu$, $\tr(E_\mu\sigma)=\tr(E_\mu\rho)$. Summing over $\mu$ and using $\sum_\mu E_\mu=\mathbb{1}$, we obtain that $\tr(\sigma)=\tr(\rho)$, and therefore, if $\rho$ is a quantum state $\sigma$ must also be a quantum state. This implies that $V$ must be a traceless Hermitian matrix, hence, $n_-(V) \geq 1$. Using the inertia additivity formula for $\sigma$ gives,
\begin{equation}
\iner(\sigma) = \iner(A)+\iner(\sigma/A).
\end{equation}
By definition, the Schur complement is
\begin{equation}
\sigma/A=C +V - B A^{-1} B^{\dagger}=\rho/A+V=V.
\end{equation}
The inertia additivity formula for  $\sigma$ thus reads,
\begin{equation}
\iner(\sigma) = \iner(A)+\iner(V).
\end{equation}
Since $A=\rho_{0,0}>0$, $n_-(\sigma) = n_-(V) \geq 1$ so $\sigma$ has at least one negative eigenvalue, in contradiction to the assumption that it is a PSD matrix. Therefore, $\sigma \ngeq 0$ and we conclude that the POVM in Eq.~\eqref{psi-complete pure} is rank-1 strictly-complete. 

A given POVM that is rank-$r$ complete is not necessarily rank-$r$ strictly-complete in the same way as the POVM in Eq.~\eqref{psi-complete pure}. For example, the bases in Eq.~\eqref{4gmb}, correspond to a rank-$1$ complete POVM, but not to a rank-$1$ strictly-complete POVM. For these bases, we can apply a similar analysis to show that there exists a quantum state $\sigma$  with $\rk(\sigma)>1$ that matches the measured elements of $\rho$.

Given this structure, we derive the necessary and sufficient condition for a rank-$r$ complete EP-POVMs to be rank-$r$ strictly-complete. Using the notation introduced above, let us choose an arbitrary principal submatrix $M\in {\cal M}$ that was used to construct $\rho$. Such a matrix has the form of Eq.~\eqref{block_mat} where $C=BA^{-1}B^\dagger$. Let $\sigma$ be a higher-rank matrix that has the same measured elements as $\rho$, and let $\tilde{M}$ be the submatrix of $\sigma$ which spans the same subspace as $M$. Since $\sigma$ has the same measured elements as $\rho$, $\tilde{M}$ must have the form
\begin{equation}\label{Mtilde}
\tilde{M} = 
\begin{pmatrix}
 A & B^{\dagger} \\
 B & \tilde{C} 
 \end{pmatrix}\equiv\begin{pmatrix}
 A & B^{\dagger} \\
 B & C + V 
 \end{pmatrix}=M+\begin{pmatrix}
 {\bf 0} & {\bf 0} \\
 {\bf 0} & V 
 \end{pmatrix}.
 \end{equation}
Then, from Eq.~\eqref{Schur_iner}, $\iner(\tilde{M}) = \iner(A) + \iner(\tilde{M}/A) = \iner(A) + \iner(V)$, since $\tilde{M}/A = M/A + V = V$. A matrix is PSD if and only if all of its principal submatrices are PSD~\cite{Zhang11}. Therefore, $\sigma \geq 0$ if and only if $\tilde{M}\geq 0$, and $\tilde{M} \geq 0$ if and only if $n_-(A) + n_-(V) = 0$. Since $\rho\geq0$, all of its principal submatrices are PSD, and in particular $A\geq0$. Therefore,  $\sigma \geq 0$ if and only if $n_-(V) = 0$. We can repeat this logic for all other submatrices $M \in \mathcal{M}$. Hence, we conclude that the measurement is rank-$r$ strictly-complete if and only if there exists at least one submatrix $M \in \mathcal{M}$ for which every $V$ that we may add (as in Eq.~\eqref{Mtilde}) has at least one negative eigenvalue. 

A sufficient condition for an EP-POVM to be rank-$r$ strictly-complete is given in the following proposition.\\
\\
{\noindent{\bf Proposition~1.} Assume that an EP-POVM is rank-$r$ complete. If its measurement outcomes determine the diagonal elements of the density matrix, then it is a rank-$r$ strictly-complete POVM.}\\
\\
{\em Proof.} Consider a Hermitian matrix $\sigma$ that has the same measurement probabilities as $\rho$, thus the same measured elements.  If we measure all diagonal elements of $\rho$ (and thus, of $\sigma$), then for any principal submatrix $\tilde{M}$ of $\sigma$, cf. Eq.~\eqref{Mtilde}, the corresponding $V$ is traceless because all the diagonal elements of $C$ are measured. Since $V$ is Hermitian and traceless it must have at least one negative eigenvalue, therefore, $\sigma$ is not PSD matrix and the POVM is rank-$r$ strictly-complete.\hfill $\square$ \\
\\
A useful corollary of this proposition is any EP-POVM that is rank-$r$ complete can be made rank-$r$ strictly-complete simply by adding POVM elements that determine the diagonal elements of the density matrix.

The framework we developed here allows us to construct rank-$r$ strictly-complete POVMs. We present two such POVMs in Appendix~\ref{app:construction} and describe the algebraic reconstruction of the rank-$r$ state. The POVMs are generalization of the POVMs by Flammia~{\em et al.}~\cite{Flammia05} and Goyeneche~{\em et al.}~\cite{Goyeneche14} from pure states to rank-$r$ states, such that the construction of the rank-$(r-1)$ strictly-complete POVM is nested in the rank-$r$ strictly-complete POVM. The usefulness of such nested POVMs is discussed in Sec.~\ref{sec:numerics}.

\subsection{Measurement of random bases}\label{ssec:bases}
We numerically study a straightforward protocol to implement strictly-complete measurements by measuring a collection of random bases. In particular, we find that measuring only few random orthonormal bases amounts to strict-completeness. Measurement of random bases have been studied in the context of compressed sensing (see, e.g., in Refs.~\cite{Kueng15,Acharya15}). However, when taking into account the positivity of density matrices, we obtain strict-completeness with fewer measurements than required for compressed sensing~\cite{Kalev15}. Therefore, strict-completeness is not equivalent to compressed sensing. While for quantum states, all compressed sensing measurements are strictly-complete~\cite{Kalev15}, not all strictly-complete measurements satisfy the conditions required for compressed sensing estimators.

We perform the numerical experiments to determine rank-$r$ strictly-complete measurement for $r=1,2,3$.  To achieve this, we take the ideal case where the measurement outcomes are known exactly and the rank of the state is fixed. We consider two types of measurements on a variety of different dimensions: (i) a set of  Haar-random orthonormal bases on unary qudit systems with dimensions $d=11, 16, 21, 31, 41$, and $51$; and (ii) a set of local Haar-random orthonormal bases on a tensor product of $n$ qubits with $n=3,4,5$, and $6$, corresponding to $d=8, 16, 32$, and $64$, respectively. For each dimension, and for each rank, we generate $25d$ Haar-random states. For each state we calculate the noiseless measurement vector, $\bm{p}$,  with an increasing number of bases.  After each new basis measurement we use the constrained least-square (LS) program, Eq.~\eqref{general_norm_positive_CS}, where $\Vert\cdot\Vert$ is the $\ell_2$-norm, to produce an estimate of the state. We emphasize that the constrained LS finds the quantum state that is the most consistent with $\bm{p}$ with no restrictions on rank. The procedure is repeated until all estimates match the states used to generate the data (up to numerical error of $10^{-5}$ in infidelity). This indicates the random bases used correspond to a rank-$r$ strictly-complete POVM.

\begin{table}[h]
\centering
\begin{tabular}{ cc|c|c|c|c|c|||c|c|c|c| }
& \multicolumn{9}{c}{\bf{Dimension}}\\ \cline{2-11}
 & \multicolumn{6}{|c|||}{Unary}  & \multicolumn{4}{c| }{Qubits}\\
 \multicolumn{1}{ c|| }{\bf{Rank}} &\bf{11} &\bf{16}&\bf{21} &\bf{31} &\bf{41}&\bf{51} &\bf{8} &\bf{16} &\bf{32}&\bf{64} \\
\hline\hline
\multicolumn{1}{ |c|| }{\bf{1}} &\multicolumn{6}{c|||}{6} & \multicolumn{4}{c|}{6}  \\\cline{1-11} 
\multicolumn{1}{ |c|| }{\bf{2}} & 7  & 8  & 8  & \multicolumn{3}{ c||| } {9}& \multicolumn{2}{ c| } {9} &  \multicolumn{2}{ c| } {10}  \\\cline{1-11}
 \multicolumn{1}{ |c|| }{\bf{3}} & 9  & 10  & 11  & 12 & 12  & 13 & 12 & \multicolumn{3}{ c| } {15}\\\cline{1-11}
\hline
\end{tabular}
\caption{{\bf Number of random orthonormal bases corresponding to strict-completeness.}  Each cell lists the minimal number of measured bases for which the infidelity was below $10^{-5}$  for each of the tested states in the given dimensions and ranks. This indicates that a measurement of only few random bases is strictly-complete POVM.}\label{tbl:noiseless}
\end{table}  

We present our findings in Table~\ref{tbl:noiseless}.  For each dimension, we also tested fewer bases than listed in the table. These bases return infidelity below $10^{-5}$ for most states but not all. For example, in the unary system with $d=21$, using the measurement record from $5$ bases we can reconstruct all states with an infidelity below the threshold except for one. The results indicate that measuring only few random bases, with weak dependence on the dimension, corresponds to a strictly-complete POVM for low-rank quantum states. Moreover, the difference between, say rank-1 and rank-2, amounts to measuring only a few more bases. This is important, as discussed below, in realistic scenarios when the state of the system is known to be close to pure. Finally, when considering local measurements on qubits, more bases are required to account for strict-completeness when compared to unary system; see for example results for $d=16$.

\section{Numerical experiments with strictly-complete POVMs}\label{sec:numerics}
To demonstrate the robustness of the estimators, we simulate a realistic scenario where the state of the system is full rank but high purity and the experimental data contains statistical noise. From Corollary~2 we expect to obtain a robust estimation of the state by solving any convex estimator of the form of Eqs.~\eqref{general_positive_CS_noisy} and~\eqref{general_norm_positive_CS_noisy}. We use three example estimators (using the MATLAB package CVX~\cite{cvx}):  \\
(i) A constrained trace-minimization program,
\begin{equation}\label{est_tr}
\hat{X} = \arg\min_X \tr(X)\;\; {\rm s.t.}\; \Vert{\cal M}[X]-\bm{f}\Vert_2\leq\epsilon\, \,  {\rm and} \, \,  X \geq 0,
\end{equation}
(ii) a constrained LS program,
\begin{equation}\label{est_ls}
\hat{X} = \arg\min_X \Vert{\cal M}[X]-\bm{f}\Vert_2\;\; {\rm s.t.}\; X \geq 0,
\end{equation}
(iii) maximum-likelihood when constraining the model to be a quantum state,
\begin{align}\label{est_lik}
\hat\rho &= \arg\min_\rho -\sum_\mu f_\mu\log(\tr(E_\mu \rho))\\\nonumber& {\rm s.t.}\; \Vert{\cal M}[\rho]-\bm{f}\Vert_2\leq\epsilon, \, \,  \rho \geq 0, \, \,  {\rm and} \, \,\tr(\rho)=1,
\end{align}
where $\epsilon$ is generated based on the variance of the multinomial distribution if the maximally mixed state was measured, $\epsilon=\sqrt{b(1-1/d)/N}$, where $b$ is the number of bases and $N$ is the number of samples per basis. In the first two programs the trace constraint is not included hence $\hat\rho=\hat{X}/\tr(\hat{X})$.

We simulate two different types of systems. We consider unary systems of qudits with dimensions $d=11, 21$, and $31$, measured with a series of Haar-random bases. Secondly, we consider a collection of $n=3,4$, and 5 qubits.  In this case we simulate measurements both with a series of Haar-random bases on each qubit and also with the rank-$r$ generalization of the measurement proposed by Goyeneche~{\em et al.}~\cite{Goyeneche14}, defined in Appendix~\ref{app:examples_bases}. For each system we generate 100 Haar-random pure-states (target states), $\{| \psi\rangle\}$, and take the state of the system to be $\sigma =(1- q) | \psi \rangle \langle \psi | + q \tau$, where $q = 10^{-3}$, and $\tau$ is a random full-rank state generated from the Hilbert-Schmidt measure. The measurement vector, $\bm{f}$, is simulated by sampling $m = 300 d$ trials from the corresponding probability distribution. For each number of measured bases, we estimate the state with the three different convex optimization programs listed above. 

\begin{figure}[t]
\centering
\includegraphics[width=\linewidth]{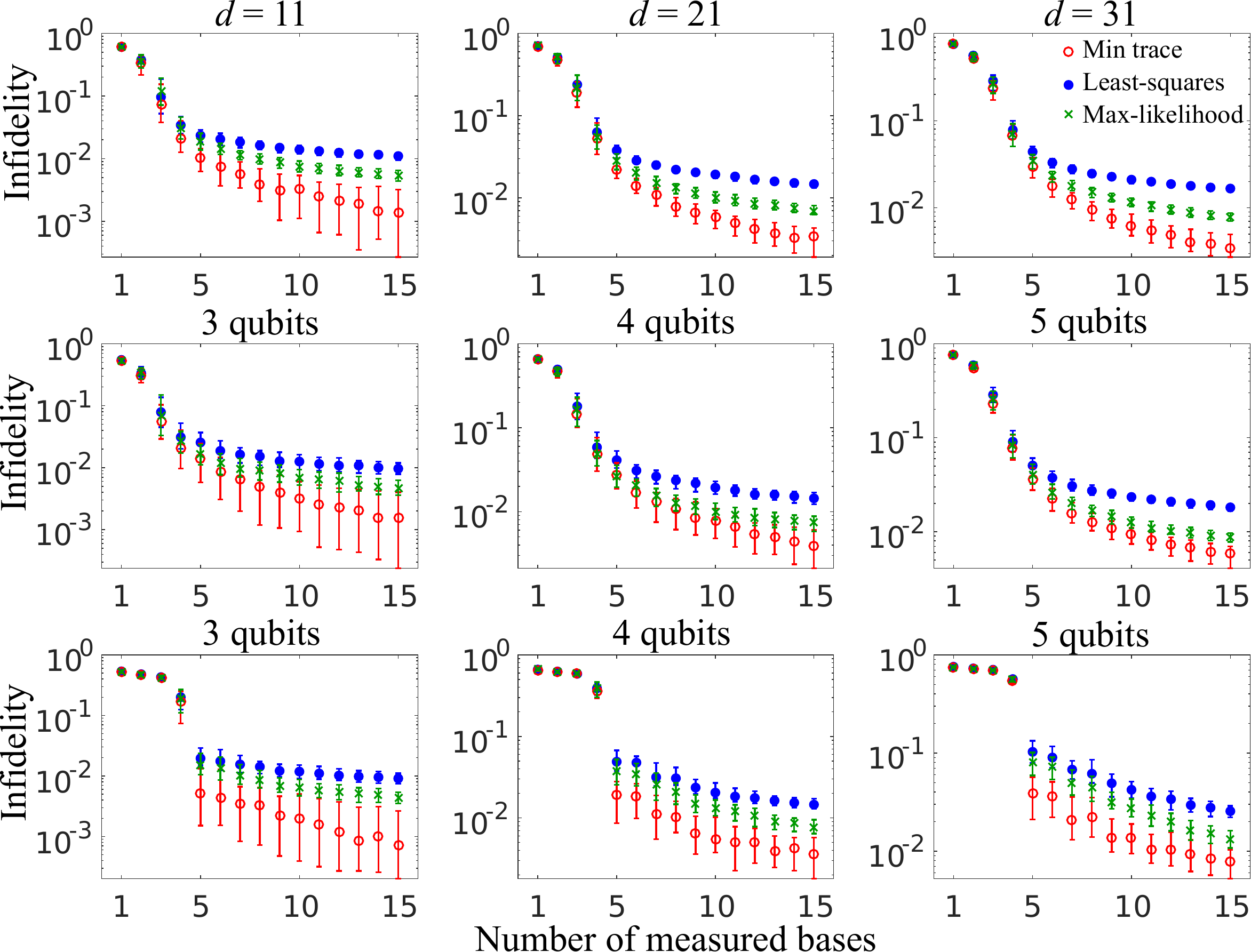}
\caption{{\bf Simulation of QST under realistic conditions.} We assume that the state of the system is a full-rank state close to a target pure state. We plot the median infidelity (on a log-scale) between the target pure state and its estimation as a function of measured bases for three different estimators Eqs.~\eqref{est_tr}-\eqref{est_lik}.  The error bars show the interquartile range (middle 50\%) of the infidelities found over 100 numerical experiments. {\bf Top row:} We simulate Haar-random measurement bases for the corresponding unary system. {\bf Middle row:} The measurement bases are randomly generated by composing tensor products of Haar-random local bases on qubit subsystems. {\bf Bottom row:} The measurement bases are the rank-$r$ Goyeneche bases, whose explicit construction is given in Appendix~\ref{app:examples_bases}. Most of the information about the state is obtained when the number of measured bases corresponds to rank-1 strictly-complete. Following Corollary~2, we obtain a robust estimation regardless of the particular program used to estimate the state.}
\label{fig:noisy}
\end{figure}

In Fig.~\ref{fig:noisy} we plot the average infidelity (over all tested states) between the target state, $ | \psi \rangle$, and its estimation, $\hat\rho$, $1-\overline{\langle\psi|\hat\rho| \psi \rangle}$. As ensured by Corollary~2, the three convex programs we used robustly estimate the state with a number of bases that correspond to rank-1 strictly-complete POVM, that is, six bases for in the case of Haar-random basis measurements, and five bases for the POVM constructed in Appendix~\ref{app:examples_bases}, based on Ref.~\cite{Goyeneche14}. Furthermore, in accordance with our findings, if one includes the measurement outcomes of only a few more bases such that the overall POVM is rank-$2$ strictly-complete, or higher, we improve the estimation accordingly.

\section{Summary and Conclusions}\label{sec:conclusions}
We have studied QST under the assumption that the state of the system is known to be close to a pure state, or more generally, to a rank $\leq r$ state. Since the set of rank $\leq r$ states is nonconvex, it is generally difficult to robustly estimate the state of the system by measuring rank-$r$ complete POVMs. We showed, however, that a robust estimation is guaranteed if the measurements are rank-$r$ strictly-complete. Such measurements  efficiently identify a low-rank state from within the set of all quantum states. The essential ingredient of strict-completeness is the positivity constraint associated with physical density matrices. Moreover, the estimation can be done by solving any convex program over the feasible set and the estimate returned is robust to errors.

Generally, it is difficult to assess if an arbitrary POVM satisfies one of these completeness relations. In this work we studied two different ways of designing strictly-complete measurements. The first was in the context of EP-POVMs, which allow for the algebraic reconstruction of a few density matrix elements. In this situation the problem of QST is reduced to density matrix completion. We developed tools to determine if a given EP-POVM is rank-$r$ complete or rank-$r$ strictly-complete based on properties of the Schur complement and matrix inertia. These tools provide a unified framework for all EP-POVMs and we used them to construct two rank-$r$ strictly-complete measurements. We also showed that a few random bases also form a strictly-complete measurement, with the number of bases required scaling weakly with the dimension.

With either of these approaches one can iteratively probe highly-pure quantum states. A rank-$1$ strictly-complete POVM could be used to produce an estimate of the dominant eigenvalue, as was shown by Goyeneche~{\em et al.}~\cite{Goyeneche14}. One can then use our generalization for rank-$r$ strictly-complete POVMs to produce more accurate estimates, when needed. For example, a rank-$2$ strictly-complete POVM, such as the ones introduced in Sec.~\ref{sec:constructions}, would produce an estimate corresponding to the state's two largest eigenvalues. One could continue to produce more accurate estimates but at some point the eigenvalues will be so small that other sources of noise will dominate. In future work we plan to explore how one can use such an iterative procedure to certify the number of dominant eigenvalues in the state without performing full quantum tomography.

\begin{acknowledgments}
This work was supported by NSF Grants PHY-1212445, PHY-1521016, and PHY-1521431.
\end{acknowledgments}

\appendix

\section{Proof of Corollary~2}\label{app:proof}
The proof of Corollary~2 uses Lemma~V.5 of~\cite{Kech15b}, restated as follows.\\
\noindent {\bf Lemma~3:} Let ${\cal E}$ be a rank-$r$ strictly-complete POVM, and let ${\bm f}= {\cal M}_{\cal E}[\sigma]+{\bm e}$ be the measurement record of some quantum state, $\sigma$. If  $\Vert{\bm f}-{\cal M}_{\cal E}[\rho_0]\Vert\leq \epsilon$ for some quantum  state $\rho_0$ with $\rk\rho_0\leq r$,   then for every PSD matrix $X$ such that $\Vert{\cal M}_{\cal E}[X]-{\bm f}\Vert\leq \epsilon$, we have $\Vert X-\rho_0\Vert\leq C_{\cal E}\epsilon$, where $C_{\cal E}$ depends only on the POVM.\\

The proof of this Lemma can be found in~\cite{Kech15b}. To prove Corollary~2, we first show that $\Vert\hat{X}-\rho_0\Vert\leq C_{\cal E}\epsilon$. The convex programs of Eqs.~\eqref{general_positive_CS_noisy} and~\eqref{general_norm_positive_CS_noisy} in the main text look for a solution that minimizes some convex function on the set  $\{\Vert{\cal M}_{\cal E}[X]-{\bm f}\Vert\leq \epsilon, X\geq0\}$. According to the Lemma,  any PSD matrix $X$ within this set satisfies $\Vert X-\rho_0\Vert\leq C_{\cal E}\epsilon$. Since the  solution $\hat{X}$ is also in that set, we obtain that $\Vert\hat{X}-\rho_0\Vert\leq C_{\cal E}\epsilon$. 

Next, we show that $\Vert\hat{X}-\sigma\Vert\leq 2C_{\cal E}\epsilon$. Since  we assume $\Vert{\bm e}\Vert\leq\epsilon$, $\sigma$ is in the set  $\{\Vert{\cal M}_{\cal E}[X]-{\bm f}\Vert\leq \epsilon, X\geq0\}$, and according to the Lemma $\Vert\sigma-\rho_0\Vert\leq C_{\cal E}\epsilon$. Therefore we have
\begin{align*}
\Vert\hat{X}-\sigma\Vert&=\Vert\hat{X}-\rho_0-\sigma+\rho_0\Vert\leq\Vert\hat{X}-\rho_0\Vert+\Vert\sigma-\rho_0\Vert\\&\leq 2C_{\cal E}\epsilon.
\end{align*}
\hfill$\square$\\
For convenience one parameter, $\epsilon$, is used to quantify the various bounds. However it is straightforward to generalize this result to the case where the bounds are quantified by different values.

\section{Construction of rank-$r$ strictly-complete POVMs}\label{app:construction}
The framework we developed in Sec.~\ref{ssec:ep} allows us to construct rank-$r$ strictly-complete POVMs. We present here two such constructions. 

\subsection{Rank-$r$ Flammia}\label{app:examples_full}
A rank-$r$ density matrix has $(2d-r)r-1$ free parameters. The first rank-$r$ strictly-complete POVM we form has $(2d-r)r+1$ elements, and is a generalization of the POVM in Eq.~\eqref{psi-complete pure}. The POVM elements are,
\begin{align}\label{psic mixed}
&E_k=a_k\ket{k}\bra{k},\;k=0,\ldots,r-1\nn
&E_{k,n}=b_k(\mathbb{1}+\ket{k}\bra{n}+\ket{n}\bra{k}),\;n=k+1,\ldots,d-1,\nn
&\widetilde{E}_{k,n}=b_k(\mathbb{1}-\ii\ket{k}\bra{n}+\ii\ket{n}\bra{k}), \;n=k+1,\ldots,d-1,\nn
&E_{(2d-r)r+1}=\mathbb{1}-\sum_{k=0}^{r}\Bigl[E_k +\sum_{n=1}^{d-1}(E_{k,n}+\widetilde{E}_{k,n})\Bigr],
\end{align}
with $a_k$ and $b_k$ chosen such that $E_{(2d-r)r+1}\geq0$. The probability $p_k=\tr(E_k\rho)$ can be used to calculate the density matrix element $\rho_{k,k}=\matele{k}{\rho}{k}$, and the probabilities $p_{k,n}=\tr(E_{k,n}\rho)$ and $\tilde{p}_{k,n}=\tr(\widetilde{E}_{k,n}\rho)$ can be used to calculate the density matrix elements $\rho_{n,k}=\matele{n}{\rho}{k}$ and $\rho_{k,n}=\matele{k}{\rho}{n}$. Thus, this is an EP-POVM which reconstruct the first $r$ rows and first $r$ columns of the density matrix. 

Given the measured elements, we can write the density matrix in block form corresponding to measured and unmeasured elements,
\begin{equation} \label{block_rho_gen}
\rho= 
\begin{pmatrix}
A &  B^\dagger\\
B & C 
\end{pmatrix},
\end{equation}
where $A$ is a $\by{r}$ submatrix and $A$, $B^\dagger$, and $B$ are composed of measured elements. Suppose that $A$ is nonsingular. Given that $\rk(\rho)=r$, using the rank additivity property of Schur complement and that $\rk(A)=r$, we obtain $\rho/A=C-BA^{-1}B^\dagger=0$. Therefore, we conclude that $C=BA^{-1}B^\dagger$.  Thus we can reconstruct the entire rank-$r$ density matrix. 

Following the arguments for the POVM in Eq.~\eqref{psi-complete pure}, it is straight forward to show that this POVM is in fact rank-$r$ strictly-complete. The failure set of this POVM corresponds to states for which $A$ is singular. The set is dense on a set of states of measure zero. 

The POVM of Eq.~\eqref{psic mixed} can alternatively be implemented as a series of $r-1$ POVMs, where the $k$th POVM, $k=0,\ldots,r-1$, has $2(d-k)$ elements, 
\begin{align}\label{psic mixed kth}
&E_k=a_k\ket{k}\bra{k},\nn
&E_{k,n}=b_k(\mathbb{1}+\ket{k}\bra{n}+\ket{n}\bra{k}),\;n=k+1,\ldots,d-1,\nn
&\widetilde{E}_{k,n}=b_k(\mathbb{1}-\ii\ket{k}\bra{n}+\ii\ket{n}\bra{k}), \;n=k+1,\ldots,d-1,\nn
&E_{2(d-k)}=\mathbb{1}-\Bigl[E_k +\sum_{n=1}^{d-1}(E_{k,n}+\widetilde{E}_{k,n})\Bigr].
\end{align}

\subsection{Rank-$r$ Goyeneche}\label{app:examples_bases}
The second rank-$r$ strictly-complete POVM we  construct corresponds to a measurement of $4r+1$ orthonormal  bases, which is a generalization of the four basis in Eq.~\eqref{4gmb}. We consider the case that the dimension of the system is a power of two. Since a measurement of $d+1$ mutually unbiased bases is fully informationally complete~\cite{Wootters89}, this construction is relevant as long as $r< d/4$.  We first assess the case of $r=1$, which is the measurement proposed by Goyeneche~{\em et al.}~\cite{Goyeneche14} but without adaptation. In this case there are five bases, the first is the computational basis, $\{\ket{k}\}$, $k=0,\ldots,d-1$ and the other four are given in Eq.~\eqref{4gmb}. Goyeneche~{\em et al.}~\cite{Goyeneche14} showed that the last four bases are rank-1 complete. Here, we show these five bases are rank-1 strictly-complete with the techniques introduced above.

We label the upper-right diagonals $0$ to $d-1$, where the $0$th diagonal is the principal diagonal and the $(d-1)$st diagonal is the upper right element. Each diagonal, except the $0$th, has a corresponding Hermitian conjugate diagonal (its corresponding lower-left diagonal). Thus, if we measure the elements on a diagonal, we also measure the elements of its Hermitian conjugate. The computational basis corresponds to measuring the $0$th diagonal. In Sec.~\ref{ssec:ep} we showed measuring the last four bases corresponds to measuring the elements on the first  diagonals. To show that the measurement of these five bases is rank-1 complete, we follow a similar strategy outlined in Sec.~\ref{ssec:ep}. First, choose the leading $\by{3}$ principal submatrix,
\begin{equation}
M_0 = 
\begin{pmatrix}
\rho_{0,0} & \rho_{0,1} & \bm{\rho_{0,2}} \\
\rho_{1,0} & \rho_{1,1} & \rho_{1,2} \\
\bm{\rho_{2,0}} & \rho_{2,1} &\rho_{2,2}  \\
\end{pmatrix},
\end{equation}
where, hereafter, the elements in bold font are the unmeasured elements. By applying a unitary transformation, which switches the first two rows and columns, we can move $M_0$ into the canonical form, 
\begin{equation}
M_0 \rightarrow UM_0U^\dagger=
\begin{pmatrix}
\rho_{1,1} &\rho_{1,0} & \rho_{1,2} \\
\rho_{0,1} & \rho_{0,0} & \bm{\rho_{0,2}} \\
 \rho_{2,1} & \bm{\rho_{2,0}} & \rho_{2,2}   \\
\end{pmatrix}.
\end{equation}
From Eq.~\eqref{Schur_rank} we can solve for the bottom $\by{2}$ block of $UM_0U^\dagger$ if $\rho_{1,1} \neq 0$. The set of states with $\rho_{1,1}= 0$ corresponds to the failure set. Note that the diagonal elements of the bottom $\by{2}$ block, $\rho_{0,0}$ and $\rho_{2,2}$, are also measured. We repeat this procedure for the set of principal $\by{3}$ submatrices, $M_{i} \in \mathcal{M}$, $i=0,\ldots,d-2$,
\begin{equation}
M_{i} = \begin{pmatrix}
\rho_{i,i} & \rho_{i,i+1} & \bm{\rho_{i,i+2}} \\
\rho_{i+1,i} & \rho_{i+1,i+1} & \rho_{i+1,i+2} \\
\bm{\rho_{i+2,i}} & \rho_{i+2,i+1} &\rho_{i+2,i+2}  \\
\end{pmatrix},
\end{equation}
For each $M_{i}$, the upper-right and the lower-left corners elements $\rho_{i,i+2}$ and $\rho_{i+2,i}$ are unmeasured. Using the same procedure as above we reconstruct these elements for all values of $i$ and thereby reconstruct the 2nd diagonals. We repeat the entire procedure again choosing a similar set of $\by{4}$ principal submatrices and reconstruct the 3rd diagonals and so on for the rest of the diagonals until all the unknown elements of the density matrix are reconstructed. Since, we  have reconstructed all diagonal elements of the density matrix and used the assumption that $\rk{(\rho)} = 1$ these five bases correspond to rank-$1$ complete POVM. The first basis measures the 0th diagonal so by Proposition~1 the measurement is rank-1 strictly-complete.

The failure set corresponding to ${\cal M}$ is when $\rho_{i,i} = 0$ for $i = 1,\ldots,d-2$. Additionally, the five bases provide another set of submatrices ${\cal M}'$ to reconstruct $\rho$. This set of submatrices results from also measuring the elements $\rho_{d-1,0}$ and $\rho_{0,d-1}$, which were not used in the construction of ${\cal M}$. The failure set for ${\cal M}'$ is the same as the failure set of ${\cal M}$ but since ${\cal M'} \neq {\cal M}$ we gain additional robustness. When we consider both sets of submatrices the total failure set is $\rho_{i,i} = 0$ and $\rho_{j,j} = 0$ for $i = 1,\ldots,d-2$ and $i \neq j \pm 1$. This is the exact same set found by Goyeneche~{\em et al.}~\cite{Goyeneche14}.

We generalize these ideas to measure a rank-$r$ state by designing $4r +1$ orthonormal bases that correspond to a rank-$r$ strictly-complete POVM. The algorithm for constructing these bases, for dimensions that are powers of two, is given in Algorithm~\ref{alg:rankr_GMB}. Technically, the algorithm produces unique bases for $r \leq d/2$ but, as mentioned before, since $d+1$ mutually unbiased bases are informationally complete, for $r \geq d/4$ one may prefer to measure the latter. The corresponding measured elements are the first $r$ diagonals of the density matrix. 

Given the first $r$ diagonals of the density matrix, we can reconstruct a state $\rho \in {\cal S}_r$ with a similar procedure as the one outlined for the five bases. First, choose the leading $\by{(r+2)}$ principle submatrix, $M_0$. The unmeasured elements in this submatrix are $\rho_{0,r+1}$ and $\rho_{r+1,0}$. By applying a unitary transformation we can bring $M_0$ into canonical form, and by using the rank condition from Eq.~\eqref{Schur_rank} we can solve for the unmeasured elements. We can repeat the procedure with the set of $\by{(r+2)}$ principle submatrices $M_i \in {\cal M}$ for for $i = 0,\ldots,d-r-1$ and 
\begin{equation}
M_{i} = \begin{pmatrix}
\rho_{i,i} & \cdots & \bm{\rho_{i,i+r+1}} \\
\vdots & \ddots & \vdots  \\
\bm{\rho_{i+r+1,i}} & \cdots &  \rho_{i+r+1,i+r+1}
\end{pmatrix}.
\end{equation}
From $M_i$ we can reconstruct the elements $\rho_{i,i+r+1}$, which form the $(r+1)$st diagonal. We then repeat this procedure choosing the set of $\by{(r+3)}$ principle submatrices to reconstruct the $(r+2)$nd diagonal and so on until all diagonals have been reconstructed. This shows the measurements are rank-$r$ complete and by Proposition~1, since we also measure the computational bases, the POVM is also rank-$r$ strictly-complete. 

The failure set corresponds to the set of states with singular $\by{r}$ principal submatrix
\begin{equation}
A_i = \begin{pmatrix}
\rho_{i+1,i+1} & \cdots & \rho_{i,i+r} \\
\vdots & \ddots & \vdots \\
\rho_{i+r,i} & \cdots & \rho_{i+r,i+r}
\end{pmatrix},
\end{equation}
for $i = 1,\ldots,d-r-1$. This procedure also has robustness to this set since, as in the case of $r = 1$, there is an additional construction ${\cal M}'$. The total failure set is then when $A_i$ is singular for $i = 0,\ldots,d-r-1$ and $A_j$ is singular for $j \neq i \pm 1$.

\begin{widetext}
\section*{}
\begin{algorithm}[H]
\caption{Construction of $4r +1$ bases that compose a rank-$r$ strictly-complete POVM}
 \label{alg:rankr_GMB}
\begin{enumerate}\label{alg:rankr_GMB}
\item{Construction of the first basis:}
\item[]{The choice of the first basis is arbitrary, we denote it by $\mathbbm{B}_0 = \{ |0 \rangle, | 1 \rangle, \ldots , | d-1 \rangle \}$. This basis defines the representation of the density matrix. Measuring this basis corresponds to the measurement of the all the elements on the 0th diagonal of $\rho$.}
\item{Construction of the other $4r$ orthonormal bases:}
\item[]{{\bf for} $k \in [1, r]$, {\bf do}}
\begin{itemize}
\item[]{Label the elements in the $k$th diagonal of the density matrix by $\rho_{m,n}$ where $m = 0,\ldots,d - 1 - k$ and $n = m + k$.}
\item[]{For each element on the $k$th and $({d-k})$th diagonal, $\rho_{m,n}$, associate two, two-dimensional, orthonormal bases,
\begin{align} \label{2dim_bases}
\mathbbm{b}^{(m,n)}_{x}=&\Bigl\{| x_{m,n}^{\pm} \rangle = \frac{1}{\sqrt{2}} \left( | m \rangle \pm | n\rangle \right)\Bigr\}, \nonumber \\
\mathbbm{b}^{(m,n)}_{y}=&\Bigl\{| y_{m,n}^{\pm} \rangle = \frac{1}{\sqrt{2}} \left( | m \rangle \pm \ii | n \rangle \right)\Bigr\},
\end{align}
for allowed values of $m$ and $n$.}
\item[]{Arrange the matrix elements of the $k$th diagonal and $({d-k})$th diagonal into a vector with $d$ elements 
\begin{equation}
\vec{v}(k) = ( \underbrace{\rho_{0,k}, \ldots, \rho_{d-1-k,d-1}}_\text{$k$th diagonal elements},\underbrace{\rho_{0,d-i},\ldots,  \rho_{k-1,d-1}}_\text{$({d-k})$th diagonal elements})\equiv(v_1(k),\ldots,v_d(k)). 
\end{equation}}
\item[]{Find the largest integer $Z$ such that $\frac{k}{2^{Z}}$ is an integer.}
\item[]{Group the elements of $\vec{v}(k)$ into two vectors, each with $d/2$ elements, by selecting $\ell = 2^Z$ elements out of $\vec{v}(k)$ in an alternative fashion,
\begin{align}
\vec{v}^{(1)}(k)  &= ( v_1, \ldots, v_\ell, v_{2\ell+1}, \ldots, v_{3\ell}, \ldots, v_{d-2\ell+1}, \ldots, v_{d-\ell} ) =(\rho_{0,i},\ldots,\rho_{\ell,i+\ell},\ldots),\nonumber \\
\vec{v}^{(2)}(k) &=( v_{\ell+1}, \ldots, v_{2\ell}, v_{3\ell+1}, \ldots, v_{4\ell}, \ldots, v_{d-\ell+1}, \ldots, v_{d})=(\rho_{\ell+1,i+\ell+1},\ldots, \rho_{2\ell,i+2\ell},\ldots)\nonumber
\end{align}}
\item[]{{\bf for} $j=1,2$ {\bf do}}
\begin{itemize}
\item[]{Each element of $\vec{v}^{(j)}(k)$ has two corresponding bases $\mathbbm{b}^{(m,n)}_{x}$ and $\mathbbm{b}^{(m,n)}_{y}$ from Eq.~\eqref{2dim_bases}.}
\item[]{Union all the two-dimensional orthonormal $x$-type bases into one basis
\begin{equation}
\mathbbm{B}^{(k;j)}_{x}=\bigcup_{\rho_{m,n}\in\vec{v}^{(j)}(k)} \mathbbm{b}^{(m,n)}_{x}.
\end{equation}
Union all the two-dimensional orthonormal $y$-type bases into one basis
\begin{equation}
\mathbbm{B}^{(k;j)}_{y}=\bigcup_{\rho_{m,n}\in\vec{v}^{(j)}(k)} \mathbbm{b}^{(m,n)}_{y}.
\end{equation}
The two bases $\mathbbm{B}^{(k;j)}_{x}$ and $\mathbbm{B}^{(k;j)}_{y}$ are orthonormal bases for the $d$-dimensional Hilbert space. }
\end{itemize}
\item[]{{\bf end for}}
\item[]{By measuring $\mathbbm{B}^{(k;j)}_{x}$ and $\mathbbm{B}^{(k;j)}_{y}$ for $j=1,2$ (four bases in total), we measure all the elements on the $k$th and $(d-k)$th off-diagonals of the density matrix.}
\end{itemize}
\item[]{{\bf end for}}
\end{enumerate}
\end{algorithm}
\end{widetext}


\begin{thebibliography}{99}
\bibitem{emerson05} J. Emerson, R. Alicki, and K. \.{Z}yczkowski, J. Opt. B {\bf 7}, S347 (2005).
\bibitem{knill08} E. Knill, D. Leibfried, R. Reichle, J. Britton, R. B. Blakestad, J. D. Jost, C. Langer, R. Ozeri, S. Seidelin, and D. J. Wineland,  Phys. Rev. A {\bf 77}, 012307 (2008).
\bibitem{magesan11} E. Magesan, J. M. Gambetta, and J. Emerson,  Phys. Rev. Lett.  {\bf 106}, 180504 (2011).
\bibitem{Flammia05} S. T. Flammia, A. Silberfarb, and C. M. Caves,  Found. Phy., {\bf 35}, 1985 (2005).
\bibitem{Finkelstein04} J. Finkelstein, Phys. Rev. A {\bf 70}, 052107 (2004).
\bibitem{gross10} D. Gross, Y.-K. Liu,  S. T. Flammia, S. Becker,  and J. Eisert, Phys. Rev. Lett. {\bf 105}, 150401 (2010).
\bibitem{liu11}  Y.-K. Liu,  Advances in Neural Information Processing Systems (NIPS) {\bf 24} 1638-1646 (2011).
\bibitem{Heinosaari13} T. Heinosaari, L. Mazzarella, and M. M. Wolf, Commun. Math. Phys. {\bf 318}, 355-374 (2013).
\bibitem{Chen13} J. Chen, H. Dawkins, Z. Ji, N. Johnston, D. Kribs, F. Shultz, and B. Zeng, Phys. Rev. A {\bf 88}, 012109 (2013).
\bibitem{Goyeneche14} D. Goyeneche, G. Ca{\~n}as, S. Etcheverry, E. S. G{\'o}mez, G. B. Xavier, G. Lima, and A. Delgado, Phys. Rev. Lett. {\bf 115}, 090401 (2015).
\bibitem{Carmeli14} C. Carmeli, T. Heinosaari, J. Schultz, and A. Toigo, J. Phys. A: Math. Theor. {\bf 47}, 075302 (2014).
\bibitem{Kalev15}A. Kalev, R. L. Kosut, and I. H. Deutsch,  {\em npj} Quant. Info. {\bf 1},  15018 (2015).
\bibitem{Carmeli15} C. Carmeli, T. Heinosaari, J. Schultz, and A. Toigo, Eur. Phys. J. D {\bf 69}, 11 (2015).
\bibitem{Kech15} M. Kech and M. M. Wolf, Preprint  arXiv:1507.00903 (2015).
\bibitem{Ma16} X. Ma, {\em et al.}, Phys. Rev. A {\bf 93}, 032140 (2016).
\bibitem{Hector15}  H. Sosa-Martinez, N. Lysne, C. H. Baldwin, A. Kalev, I. H. Deutsch, P. S. Jessen, in preparation (2015).
\bibitem{Kech15b} M. Kech, Preprint   arXiv:1511.01433 (2015).
\bibitem{Hradil97}  Z. Hradil,  {\em Phys. Rev. A} {\bf 55}, R1561 (1997).
\bibitem{Bakonyi11} M. Bakonyi, and J. J. Woerdeman, ``Matrix Completions, Moments, and Sums of Hermitian Squares'' Princeton University Press (2011).
\bibitem{Haynsworth68} E. V. Haynsworth,  Linear Algebra Appl. {\bf 1}, 73-81 (1968).
\bibitem{Zhang11}
See, e.g., the book by F. Zhang, Matrix Theory: Basic Results and Techniques. New York: Springer, (2011).
\bibitem{footnote_rank}%
Rank-$r$ complete POVMs can also completely specify states whose rank is smaller than $r$. If the state of the system has rank smaller than $r$,  then any $\by{r}$ principal submatrix of $\rho$ is singular. In this case, we chose a the largest principal matrix, $A$, which is nonsingular. Its dimension corresponds to the rank of the density matrix. 
\bibitem{footnote_tr}%
More generally, a POVM may give more information than certain matrix elements. This information could, and sometimes should, also be taken into account for the reconstruction of the density matrix. This case is beyond the scope of the current paper. 
\bibitem{Kueng15} R. Kueng, International Conference on Sampling Theory and Applications  (SampTA) IEEE, 402-406 (2015).
\bibitem{Acharya15} A. Acharya, T. Kypraios, and M. Guta, , Preprint  arXiv:  1510.03229 (2015).
\bibitem{cvx} Software for disciplined convex programming can found
at http://cvxr.com/. 
\bibitem{Wootters89} W. K. Wootters and B. D. Fields, Ann. Phys.  {\bf 191}, 363-381 (1989).
\end{thebibliography}
\end{document}